\documentclass[prd,twocolumn,showpacs,amsmath,amssymb,superscriptaddress,floatfix,nofootinbib]{revtex4}
\usepackage{amsfonts,color,graphicx,epsfig,amsmath,multirow,mathtools,amsmath}
\usepackage{subfigure,bm,slashed,lineno}
\usepackage[dvipdfm,colorlinks=true,citecolor=blue,linkcolor=blue,urlcolor=blue]{hyperref}

\begin{document}

\title{The asymmetries of the $B \to K^* \mu^+ \mu^-$ decay and the search of new physics beyond the Standard Model}

\author{Hai-Bing Fu}
\email{fuhb@cqu.edu.cn}
\address{Department of Physics, Guizhou Minzu University, Guiyang 550025, P. R. China}

\author{Xing-Gang Wu}
\email{wuxg@cqu.edu.cn}
\address{Department of Physics, Chongqing University, Chongqing 401331, P. R. China}

\author{Wei Cheng}
\email{chengw@cqu.edc.cn}
\address{Department of Physics, Chongqing University, Chongqing 401331, P. R. China}

\author{Tao Zhong}
\email{zhongtao@htu.edu.cn}
\address{Physics Department, Henan Normal University, Xinxiang 453007, P. R. China}

\author{Zhan Sun}
\email{zhansun@cqu.edu.cn}
\address{Department of Physics, Guizhou Minzu University, Guiyang 550025, P. R. China}

\date{\today}

\begin{abstract}
In this paper, we compute the forward-backward asymmetry and the isospin asymmetry of the $B \to K^* \mu^+ \mu^-$ decay. The $B \to K^*$ transition form factors (TFFs) are key components of the decay. To achieve a more accurate QCD prediction, we adopt a chiral correlator for calculating the QCD light-cone sum rules for those TFFs with the purpose of suppressing the uncertain high-twist distribution amplitudes. Our predictions show that the asymmetries under the Standard Model and the Minimal Supersymmetric Standard Model with minimal flavor violation are close in shape for $q^2 \ge 6~{\rm GeV}^2$ and are consistent with the Belle, LHCb and CDF data within errors. When $q^2 < 2~{\rm GeV}^2$, their predictions behave quite differently. Thus a careful study on the $B \to K^* \mu^+ \mu^-$ decay within small $q^2$-region could be helpful for searching new physics beyond the Standard Model. As a further application, we also apply the $B \to K^*$ TFFs to the branching ratio and longitudinal polarization fraction of the $B\to K^*\nu\bar\nu$ decay within different models.
\end{abstract}



\maketitle

\section{Introduction}

Processes involving flavor changing neutral current (FCNC) provide important platforms for testing the Standard Model (SM) and for searching of new physics beyond the SM. Among them, the $B$-meson exclusive decays, such as the $B\to K^*\mu^+\mu^-$ with the cascade decay $K^*\to K\pi$, is important. This is because the measurements of their four-body final state angular distributions provide abundant information on probing and discriminating different scenarios of new physics.

The $B$-meson exclusive decay requires a proper factorization of the long-distance and the short-distance physics, which could generally be distinguished by the heavy quark mass $m_b$ emerged in the hadronic matrix elements. By further taking the heavy-quark limit, $m_b\to \infty$, the hadronic amplitudes arising from the hard gluon exchanges can be factorized into the perturbative scattering kernels and the nonperturbative but universal hadronic quantities. This treatment has been successfully introduced in dealing with the nonleptonic $B$-meson decays, the heavy-to-light transition form factors, and the radiative $B$-meson decays~\cite{Beneke:2009az, Beneke:2001at, Kagan:2001zk, DescotesGenon:2002mw}.

In the paper, we shall focus on the forward-backward and the isospin asymmetries of the $B\to K^*\mu^+\mu^-$ exclusive decay, which are sensitive to the Wilson coefficients and could be used to test the new physics scenario beyond SM. The new physics part of Wilson coefficients are model dependent, which have been dealed with various methods~\cite{Lyon:2013gba, Lyon:2014hpa, Alok:2009tz, Alok:2010zd, Alok:2011gv}. According to the minimal supersymmetric standard model (MSSM) with minimal flavour violation (MFV), all flavour transitions occur only in the charged-current sector and are determined by the known CKM mixing angles. This idea is also adopted by several theoretical schemes in which the communication of the supersymmetry breaking to the observable particles occurs via flavour-independent interactions. In many of those schemes the departure from the MFV hypothesis is rather small~\cite{Dine:1993yw}. To increase the predictivity of the MSSM with MFV, one may follow several restrictions which state that all supersymmetric particles except for charginos, sneutrinos and charged Higgs fields are about $1$ TeV; heavy particles shall be integrated out, resulting in a ``low-energy'' effective theory in terms of light SUSY and SM particles; the weak effective hamiltonian includes only the SM operators and the down-squark sector including a flavour diagonal mass matrix~\cite{Feldmann:2002iw}. In the paper, we shall adopt the Wilson coefficients under the SUSY MFV model as an explanation of how the new physics terms could affect the SM predictions.

Furthermore, large uncertainties in predicting the $B\to K^*\mu^+\mu^-$ decay come from the nonperturbative quantities, namely the $B \to K^*$ transition form factors (TFFs). Those TFFs have been studied within various approaches such as the relativistic quark model~\cite{Faessler:2002ut, Ebert:2010dv}, the light-cone sum rules (LCSR)~\cite{Ball:1998kk, Ball:2004rg, Khodjamirian:2006st, Khodjamirian:2010vf, Ball:2005vx,Ali:1999mm}, the lattice QCD~\cite{Becirevic:2006nm, Liu:2011raa, Horgan:2013hoa}, and etc.. The LCSR predictions are reliable from the hard region around the large recoil point to the soft contribution below $m_b^2-2m_b\chi$ ($\chi\sim 500$ MeV is the typical hadronic scale of the decay), thus it provides an important bridge for connecting the results of various approaches and for comparing with the data.

The LCSR is based on the operator product expansion (OPE), which parameterizes the nonperturbative dynamics into light-cone distribution amplitudes (LCDAs). To compare with that of the usually considered pseudo-scalar LCDA, the vector meson's LCDA has much complex twist structures. Even though the high-twist LCDAs are generally power suppressed, their contributions are sizable, especially in specific kinematic region. The inaccurateness of high-twist LCDAs then lead to important systematic errors for the LCSR predictions. A practical way to suppress the uncertainties from those uncertain high-twist LCDAs is to take a proper LCSR correlator. For example, the contributions from the high-twist LCDAs can be highly suppressed by using chiral correlators~\cite{Fu:2014uea, Fu:2014pba, Cheng:2017bzz}. As an application of those more accurate TFFs, we shall recalculate the forward-backward and isospin asymmetries of the $B\to K^*\mu^+\mu^-$ decay and also the branching ratio and longitudinal polarization fraction of the $B\to K^*\nu\bar\nu$ decay.

The remaining parts of the paper are organized as follows. In Sec.II, we describe our calculation technology for deriving the forward-backward and isospin asymmetries. In Sec.III, we present numerical results and discussions on the TFFs and the asymmetries of the $B\to K^*\mu^+\mu^-$ decay within the SM and the SUSY with MFV. And we also present the results for the branching ratio and longitudinal polarization fraction of the $B\to K^*\nu\bar\nu$ decay within different models in Sec.III. Sec.IV is reserved for a summary.

\section{Calculation technology}

Within the SM, the $B\to K^*\mu^+\mu^-$ decay is induced by a set of operators $\mathcal{O}_i$ appearing in the weak effective hamiltonian~\cite{Buchalla:1995vs},
\begin{equation}
H_{\rm eff} = \frac{G_F}{\sqrt2} \left\{ \sum\limits_{i=1}^2  (\lambda_u  C_i {\cal O}_i^u + \lambda_c C_i {\cal O}_i^c)  - \lambda_t \sum\limits_{i=3}^{10}  C_i {\cal O}_i \right\}, \label{heff}
\end{equation}
where $\lambda_q = V_{qs}^* V_{qb}$, and the Wilson coefficients $C_i$ are perturbatively calculable, whose values shall be alternated when new particles beyond the SM are included.

The differential decay width of $B \to K^* \mu^+ \mu^-$ over the squared transition momentum ($q^2$) and the angle ($\theta$) takes the form~\cite{Beneke:2001at},
\begin{eqnarray}
&&\frac{d^2\Gamma}{dq^2d\cos\theta} = \frac{G_F^2 |V_{ts}^*V_{tb}|^2}{128\pi^3}m_B^3\lambda^{3/2}(q^2) \left(\frac{\alpha_{\rm em}}{4\pi}\right)^2
\nonumber\\
&&\qquad \times\Bigg\{2s(1+\cos^2\theta) \xi_\perp(q^2)^2 \bigg(|\mathcal{C}_9^{(0)\perp}(q^2)|^2
\nonumber\\
&&\qquad+ |C_{10}(\mu_b)|^2\bigg)+(1-\cos^2\theta) \left(\frac{E_{K^*} \xi_\parallel(q^2)}{m_{K^*}}\right)^2
\nonumber\\
&&\qquad\times \bigg(|\mathcal{C}_9^{(0)\|}(q^2)|^2 + |C_{10}(\mu_b)|^2\Delta_\parallel(q^2)^2\bigg)
\nonumber\\
&&\qquad-8s\cos\theta \xi_\perp(q^2)^2 \Re e [\mathcal{C}_9^{(0)\perp}(q^2)] C_{10}(\mu_b) \Bigg\}, \label{dGamma}
\end{eqnarray}
where $\theta$ is the angle between the positively charged muon and the $B$ meson in the center-of-mass frame of the muon pair. The phase-space factor $\lambda(q^2) = (1-s)^2 - 2r(1+ s)+r^2$ with $s = q^2/m_B^2$ and $r = m_{K^*}^2/m_B^2$. $q^2$ is the invariant mass of the muon pair and $\alpha_{\rm em} = g^2_{\rm em}/(4\pi)$ is the fine-structure constant. $\xi_\lambda(q^2)$ with $\lambda=(\|,\bot)$ are transverse and longitudinal $B\to K^*$ TFFs. The first two terms with angular dependence $(1\pm \cos^2\theta)$ correspond to the transversely and longitudinally polarized $K^*$-meson, respectively. The third term generates the forward-backward asymmetry with respect to the plane perpendicular to the muon momentum in the center-of-mass frame of the muon pair. The factor $\Delta_\|(q^2)$ takes the form
\begin{eqnarray}
\Delta_\| (q^2) &=& 1 + 2 C_F \bigg(L-1-\frac{\pi^2 m_{K^*}f_B f_{K^*}^\|} {{\cal N}_c m_B E_{K^*}^3 \xi_\|(q^2) } q^2 \nonumber\\
&&\quad \times \Lambda_{B,+ }^{-1}(q^2)\int_0^1dx\frac{\phi_{2;K^*}^\|(x)}{\bar x} \bigg) a_s (\mu_b),  \label{Delta}
\end{eqnarray}
where $L= -(m_b^2-q^2)\ln(1-q^2/m_b^2)/{q^2}$, $\bar x=1-x$, $a_{s}(\mu_b) = \alpha_s(\mu_b)/(4\pi)$, and $E_{K^*}=(m_B^2-q^2)/(2m_B)$. $\Lambda_{B,+ }^{-1}(q^2)$ and $\Lambda_{B,- }^{-1}(q^2)$ are inverse moments of the $B$-meson LCDAs.

Using Eq.(\ref{dGamma}), the differential forward-backward asymmetry takes the form
\begin{eqnarray}
\frac{dA_{\rm FB}}{dq^2} &=& \frac{\int_0^1 d\cos\theta \frac{d^2\Gamma}{dq^2 d\cos\theta}- \int_{-1}^0 d\cos\theta \frac{d^2\Gamma}{dq^2 d\cos\theta}} {d\Gamma/dq^2},
\label{dAFB}
\end{eqnarray}
and the differential isospin asymmetry takes the form
\begin{eqnarray}
&&\frac{dA_I(q^2)}{dq^2}= \frac{\Re e [b_d^{\perp}(q^2)-b_u^{\perp}(q^2)] ~|\mathcal{C}_9^{(0)\perp}(q^2)|^2}{|C_{10}(\mu_b)|^2 +|\mathcal{C}_9^{(0)\perp}(q^2)|^2 } \nonumber\\
&&\times \bigg[ 1+ \frac{E_{K^*}^2 \Re e [b_d^\|(q^2)-b_u^\|(q^2)] |\mathcal{C}_9^{(0)\parallel}(q^2)|^2 \xi_{\|}^2(q^2)}{4 s  m_{K^*}^2 \Re e [b_d^{\perp}(q^2)-b_u^{\perp}(q^2)]|\mathcal{C}_9^{(0)\perp}(q^2)|^2 \xi_{\perp}^2(q^2) } \bigg]
\nonumber\\
&&\times \left[1 + \frac{E_{K^*}^2 }{4 s  m_{K^*}^2} \frac{|\mathcal{C}_9^{(0)\parallel}(q^2)|^2 +|C_{10}(\mu_b)|^2 }{|\mathcal{C}_9^{(0)\perp}(q^2)|^2 + |C_{10}(\mu_b)|^2}\frac{\xi_{\parallel}^2(q^2)} {\xi_{\perp}^2(q^2)}\right]^{-1},
\label{isoasy-th}
\end{eqnarray}
where
\begin{eqnarray}
b_q^{\perp}(q^2)&=&\frac{24 \pi^2 m_B f_B e_q}{q^2 \xi_{\perp}(q^2) \mathcal{C}_9^{(0)\perp}(q^2)} \bigg[  \frac{f_{K^*}^\perp}{m_b} K_1^{\perp} (q^2) + \frac{f_{K^*}^\| m_{K^*}}{6 \bar s m_B}\nonumber\\
&& \times \Lambda_{B,+}^{-1}(q^2)K_2^{\perp} (q^2)\bigg]  \label{bperpq2}
\end{eqnarray}
and
\begin{eqnarray}
b_q^{\parallel}(q^2)&=&\frac{8 \pi^2 m_{K^*} f_B f_{K^*}^\| e_q}{m_B E_{K^*} \xi_{\parallel}(q^2) \mathcal{C}_9^{(0)\parallel}(q^2) } \Lambda_{B,-}^{-1}(q^2)  K_1^{\parallel}(q^2).   \label{bparallelq2}
\end{eqnarray}
Here $e_q$ is the electric charge of the spectator quark, $e_u=2/3$ and $e_d=-1/3$. The SM Wilson coefficients $C_9^{(0)\lambda}(q^2)$ can be read from Refs.\cite{Feldmann:2002iw, Ahmady:2014cpa}. The expressions of $K^\bot_{1,2}$ and $K_1^\|$ up to subleading $\Lambda_h/m_B$ expansion can be found in Ref.\cite{Kagan:2001zk, Beneke:2001at}, whose effects are small for $b_q^{\|}(q^2)$ but are sizable for $b_q^{\bot}(q^2)$. As a cross-check, by taking the limit $q^2 \to 0$, due to fact that the photon pole dominates the $C_9^{(0) \perp}$-coefficient, we have $A_I(0)=\Re e [b_d^{\perp}(0)-b_u^{\perp}(0)]$, which rightly equals to the isospin asymmetry of the $B \to K^* \gamma$ decay.

The expressions of the TFFs $\xi_{\lambda}(q^2)$ can be related to the usually defined $B \to K^*$ TFFs $A_{1,2}(q^2)$ and $T_{1}(q^2)$ via the following way~\cite{Ahmady:2014cpa}
\begin{eqnarray}
\xi_{\bot}(q^2) &=& T_1(q^2), \label{rela1} \\
\xi_\| (q^2)  &=& \frac{m_B + m_{K^*}}{2E_{K^*}} A_1(q^2) - \frac{m_B - m_{K^*}}{m_B} A_2(q^2). \label{rela2}
\end{eqnarray}
As mentioned in the Introduction, we adopt the expressions of the TFFs $A_{1,2}(q^2)$ and $T_{1}(q^2)$ that have been derived under the LCSR approach by using a right-handed chiral correlator~\cite{Fu:2014uea, Cheng:2017bzz} to get the final LCSRs for $\xi_{\bot}(q^2)$ and $\xi_\| (q^2)$, which take the form
\begin{widetext}
\begin{eqnarray}
\xi_\bot (q^2) &=& \frac{m_b^2 m_{K^*}^2 f_{K^*}^\bot}{m_B^2 f_B} \bigg\{ \int_0^1 \frac{du}{u} e^{\frac{m_B^2 - s(u)}{M^2}} \bigg[ \frac{1}{m_{K^*}^2}\Theta (c(u,s_0)) \phi_{2;K^*}^\bot (u,\mu) - \frac{m_b^2}{4 u^2 M^4} \widetilde{\widetilde \Theta}(c(u,s_0))\phi_{4;K^*}^\bot (u) - \frac{2}{uM^2} \nonumber\\
 &&  \times \widetilde\Theta (c(u,s_0)) I_L(u) - \widetilde\Theta (c(u,s_0)) \frac{H_3(u)}{M^2}  \bigg] + \int \mathcal D \alpha_i \int_0^1dv e^{\frac{m_B^2 - s(X)}{M^2}} \frac{5}{4 X^2 M^2} \widetilde \Theta (c(X,s_0)) \Psi_{4;K^*}^\bot (\underline \alpha) \bigg\},\label{xi_bot}
\end{eqnarray}
\begin{eqnarray}
\raggedright
\xi_\|(q^2) &=& \frac{m_b m_{K^*}^2 f_{K^*}^\| }{m_B f_B} \bigg\{ \int_0^1 \frac{du}{u}~ e^{\frac{m_B^2 - s(u)}{M^2}} \bigg\{ \frac{\mathcal{C} m_B^4 - (\mathcal{C}q^2 + u) m_B^2 + u m_{K^*}^2}{{u m_B^2m_{K^*}^2}}~\Theta (c(u,s_0)) ~ \phi_{2;K^*}^\perp(u)+ \bigg[(m_B^2 - q^2)
\nonumber\\
&& \Theta (c(u,s_0)) + \frac{m_B^2 - m_{K^*}^2}{m_B^2 M^2}\widetilde\Theta (c(u,s_0))\bigg] ~\psi_{3;{K^*}}^\|(u) + \frac{1}{4u}\bigg\{ (m_B^2 - q^2)\Theta (c(u,s_0))+\bigg[[u-(m_B^2 - q^2)(\mathcal{C}
\nonumber\\
&& - 2 m_b^2)]\frac1{uM^2}-\frac{m_{K^*}^2}{m_B^2 M^2}\bigg]
\widetilde\Theta(c(u,s_0)) + \frac{m_b^2[u(m_B^2 - m_{K^*}^2) - \mathcal{C} m_B^2(m_B^2 - q^2)]}{u^2 m_B^2 M^4} \widetilde{\widetilde \Theta}(c(u,s_0)) \bigg\} ~\phi_{4;K^*}^\bot (u)
\nonumber\\
&&  + \frac{2}{u} \bigg[(m_B^2 - q^2) \Theta(c(u,s_0)) + \frac{u(m_B^2 - m_{K^*}^2) - \mathcal{C} m_B^2 (m_B^2 - q^2)} {u m_B^2 M^2} \widetilde\Theta(c(u,s_0)) - \frac{m_B^2 - m_{K^*}^2}{ u m_B^2 M^4}(\mathcal{C} - 2 m_b^2)
\nonumber\\
&&\times \widetilde{\widetilde\Theta}(c(u,s_0))\bigg]~ I_L(u)- \bigg[(m_B^2 - q^2)~ \Theta(c(u,s_0)) - \frac{1}{u m_B^2 M^2} ~[u(m_B^2 - m_{K^*}^2)- 2m_b^2~m_B^2~(m_B^2 - q^2)]
\nonumber\\
&&\times \widetilde\Theta(c(u,s_0))\bigg] H_3(u)\bigg\}  + \int \mathcal{D} \alpha_i \int_0^1 dv ~e^{\frac{m_B^2 - s(X)}{M^2}} ~ \frac{1}{2 X^2 m_B^2  M^2}~ \bigg\{~\Big[\frac{m_B^2}{X}(m_B^2 - q^2)(\underline{\mathcal{C}} - X M^2)- m_B^2
\nonumber\\
&&  + m_{K^*}^2\Big]\left[(4v - 1)\Psi_{4;K^*}^\bot (\underline \alpha) - \widetilde \Psi_{4;K^*}^\bot (\underline \alpha)\right] - (m_B^2 - m_{K^*}^2) 4v \Phi_{3;K^*}^\bot (\underline \alpha)\bigg\},\label{xi_parallel}
\end{eqnarray}
\end{widetext}
where $\mathcal{C}=m_b^2+u^2 m_{K^*}^2-q^2$, $s(\varrho)=[m_b^2-\bar \varrho(q^2-\varrho m_{K^*}^2)]/ \varrho$ with $\varrho=(u, X)$ and $X=\alpha_1+\alpha_3$, $c(\varrho,s_0)=\varrho s_0- m_b^2 +\bar\varrho q^2 -\varrho\bar\varrho m_{K^*}^2$. $\Theta(c(\varrho,s_0))$ is the usual step function, $\widetilde\Theta(c(\varrho,s_0))$ and $\widetilde{\widetilde\Theta}(c(\varrho,s_0))$ are step functions with surface terms which are defined in Ref.\cite{Fu:2014uea}.

Those two formulas show that the LCSRs for $\xi_{\bot}(q^2)$ and $\xi_\| (q^2)$ are free of contributions from most of the high-twist LCDAs, and the remaining high-twist ones are generally suppressed by $\delta^2\sim (m_K^*/m_b)^2\sim 0.03$ to compare with the leading-twist terms; thus uncertainties from the high-twist LCDAs themselves are effectively suppressed and a more accurate prediction for the TFFs $\xi_{\bot}(q^2)$ and $\xi_\| (q^2)$ can be achieved.

\section{Numerical results and discussions}
\label{Num}

The $K^*$-meson transverse decay constant $f_{K^*}^\bot$ and the $B$-meson decay constant $f_B$ are taken as, $f_{K^*}^\bot=0.185(9)~{\rm GeV}$ and $f_{K^*}^\|=0.220(5)~{\rm GeV}$~\cite{Ball:2007zt} and $f_B=0.160\pm0.019~{\rm GeV}$~\cite{Fu:2014pba}. We set the $b$-quark pole mass $m_b=4.80\pm0.05\;{\rm GeV}$, the $K^*$-meson mass ${m_{K^*}} = 0.892$ GeV, and the $B$-meson mass $m_B = 5.279$ GeV~\cite{Agashe:2014kda}.

\subsection{The $B\to K^*$ TFFs $\xi_{\lambda}(q^2)$}

The input parameters for the TFFs $\xi_{\|,\perp}(q^2)$ are taken to be the same as the ones used by Refs.\cite{Fu:2014uea, Cheng:2017bzz}. For example, we adopt the same $K^*$-meson leading-twist LCDA $\phi_{2;K^* }^\lambda$ of Refs.\cite{Fu:2014uea, Cheng:2017bzz}to do the discussion,
\begin{eqnarray}
&& \phi_{2;K^*}^{\lambda}(x, \mu_0) = \frac{A_{2;K^*}^\lambda \sqrt{3 x \bar x} {\rm Y}} {8 \pi^{3/2} \tilde f_{K^*}^\lambda b_{2;K^*}^\lambda } [1 + B_{2;K^*}^\lambda C_1^{3/2}(\xi ) \nonumber\\
&& \quad + C_{2;K^*}^\lambda C_2^{3/2}(\xi )] \exp\left[ - b_{2;K^*}^{\lambda 2} \frac{ \bar x m_s^2 + x m_q^2 - {\rm Y}^2 }{x\bar x} \right]  \nonumber\\
&&\quad \times\left[ {\rm Erf} \bigg( b_{2;K^*}^\lambda \sqrt{\frac{\mu _0^2 + {\rm Y}^2} {x\bar x}}  \bigg) - {\rm Erf}\left( b_{2;K^*}^\lambda \sqrt{\frac{{\rm Y}^2}{x\bar x} } \right) \right],\nonumber\\ \label{DA_WH}
\end{eqnarray}
where $\mu_0\sim 1$ GeV is the factorization scale, $\textrm{Erf}(x) =\frac{2}{\sqrt{\pi}}\int^x_0 e^{-t^2} dt$, $\bar{x}=1-x$ and ${\rm Y}=\bar x m_s + x m_q$. $x$ is the momentum fraction of the $s$-quark over the $K^*$-meson. The constituent quark masses are taken as $m_q\simeq300$ MeV and $m_s\simeq450$ MeV. We adopt four constraints to set the parameters of the LCDA, i.e. the normalization condition, $\langle {\bf k}_\bot^2 \rangle_{2;K^*}^{1/2}=0.37(2)~{\rm GeV}$~\cite{Fu:2014uea}, and the two Gegenbauer moments $a_1^\bot=0.04(3)$ and $a_2^\bot=0.10(8)$ ($a_1^\|=0.03(2)$ and $a_2^\|=0.11(9)$)~\cite{Ball:2007zt}.

\begin{table}[htb]
\centering
\begin{tabular}{ccc}
\hline
                            & $\xi_\|(0)$ & $\xi_\bot(0)$ \\ \hline
Our prediction          & $0.129^{+0.006}_{-0.009}$ & $0.351^{+0.036}_{-0.035}$  \\
Ref.\cite{Ball:2004rg}       & $0.126(11)$           & $0.333(28)$           \\
Ref.\cite{Altmannshofer:2008dz}         & $0.118(8)$           & $0.266(32)$           \\
Ref.\cite{Ahmady:2014cpa}       & $0.076$                   & $0.245$                   \\
\hline
\end{tabular}
\caption{The $B\to K^*$ TFFs $\xi_\lambda(q^2)$ at the large recoil point $q^2=0$. The errors are squared average of all mentioned error sources, where the LCSR predictions of Refs.\cite{Ball:2004rg, Altmannshofer:2008dz, Ahmady:2014cpa} are presented as a comparison.  }
\label{SFF:q20}
\end{table}

We adopt the usual criteria to set the LCSR parameters, the Borel window and the continuum threshold $s_0$, of the $B\to K^*$ TFFs: I) The continuum contribution is required to be less than $30\%$ of the total LCSR, and all high-twist DAs' contributions are suppressed to be less than 15\% of the total LCSR; II) The derivatives of the LCSRs over $(-1/M^2)$ give the LCSRs for $m_B$, and for self-consistency, we require all the predicted $B$-meson masses to be full-filled in comparing with the experimental one, e.g. ${|m_B^{\rm LCSR}- m_B^{\rm exp}|}/{m_B^{\rm exp}}\leq 0.1\%$.

We present the $B\to K^*$ TFFs at the large recoil point $q^2=0~{\rm GeV}^2$ in Table \ref{SFF:q20}, where the LCSR predictions of Refs.\cite{Ball:2004rg, Altmannshofer:2008dz, Ahmady:2014cpa} are presented as a comparison. The LCSRs of Refs.\cite{Ball:2004rg, Altmannshofer:2008dz} are derived by using the usual correlator, in which all twist-2, 3, 4 LCDAs are in the LCSRs. Table~\ref{SFF:q20} shows that the LCSRs under different choice of correlators are consistent with each other within errors, indicating the LCSRs are independent to the choice of correlators. A detailed discussion of the consistency of the LCSRs under different choice of correlators can be found in Ref.\cite{Cheng:2017bzz}. The differences among different LCSRs are mainly caused by different choice of the dominant leading-twist $K^*$ LCDA. For example, the use of AdS/QCD holographic leading-twist LCDA leads to a much smaller $\xi_\|(0)$~\cite{Ahmady:2014cpa}.

The contribution from the leading-twist LCDA $\phi_{2;{K^*}}^\bot$ has been amplified by using the chiral correlator, thus the systematic errors from the the $\phi_{2;{K^*}}^\bot$ parameters shall be amplified. This leads to slightly larger error than those LCSRs for usual correlator. By comparing with the data, this fact can be inversely adopted to achieve a better constraint on $\phi_{2;{K^*}}^\bot$. The high-twist terms for the LCSRs~\cite{Ball:2004rg, Altmannshofer:2008dz} follow the $\delta$-power counting rule, which could be large for $\delta^1$ twist-3 terms. By using the chiral correlator, the high-twist LCDAs' contributions are greatly suppressed due to chiral suppression, thus their own uncertainties to the LCSR can be safely neglected and the accuracy of the LCSRs can be greatly improved. For example, we find that the contributions from the twist-3 LCDA $\Phi_{3;K^*}^\|$ and the twist-4 LCDA $\Psi_{4;K^*}^\bot$ provide less than $0.1\%$ of the total LCSRs.

\begin{figure}[htb]
\includegraphics[width=0.45\textwidth]{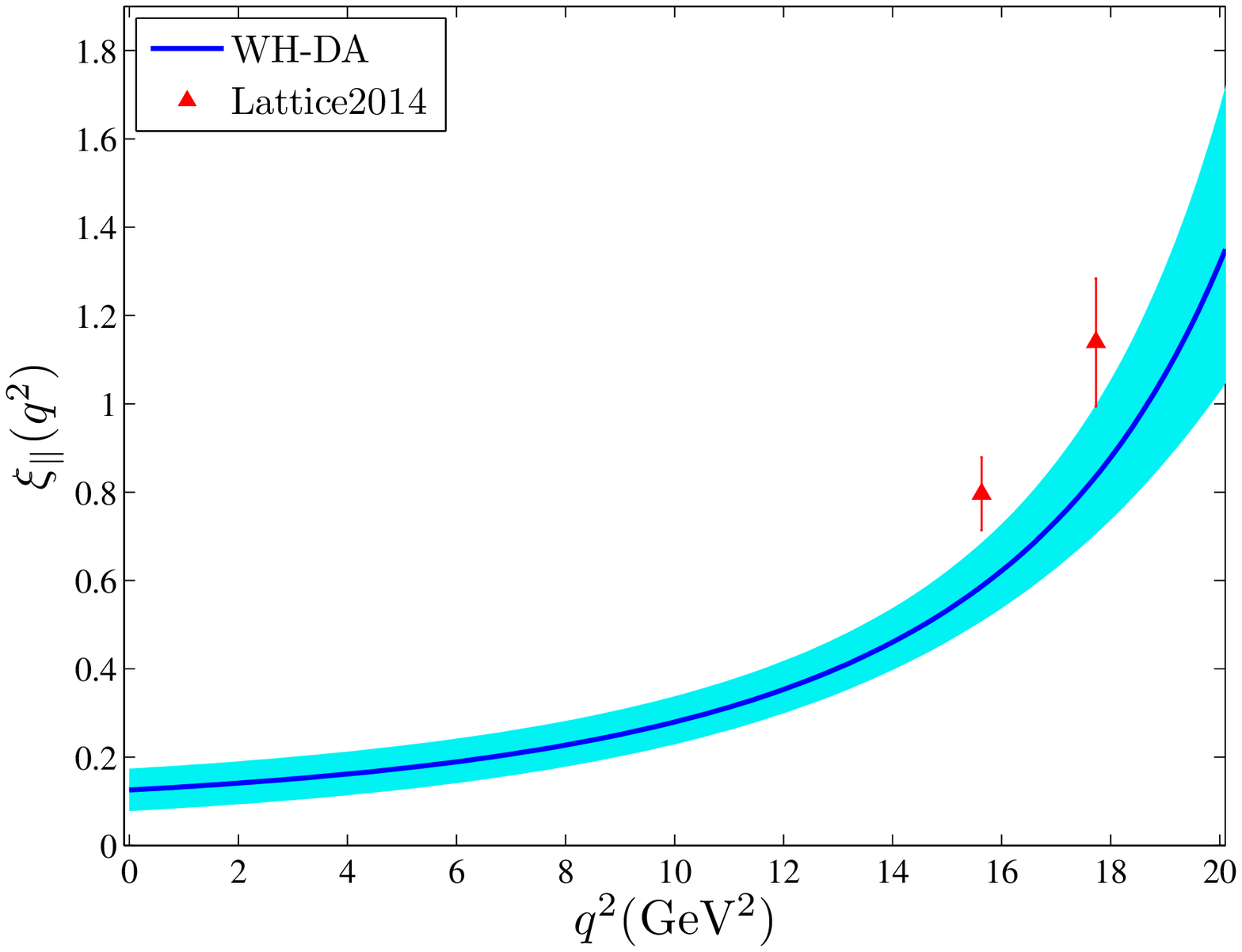}
\includegraphics[width=0.45\textwidth]{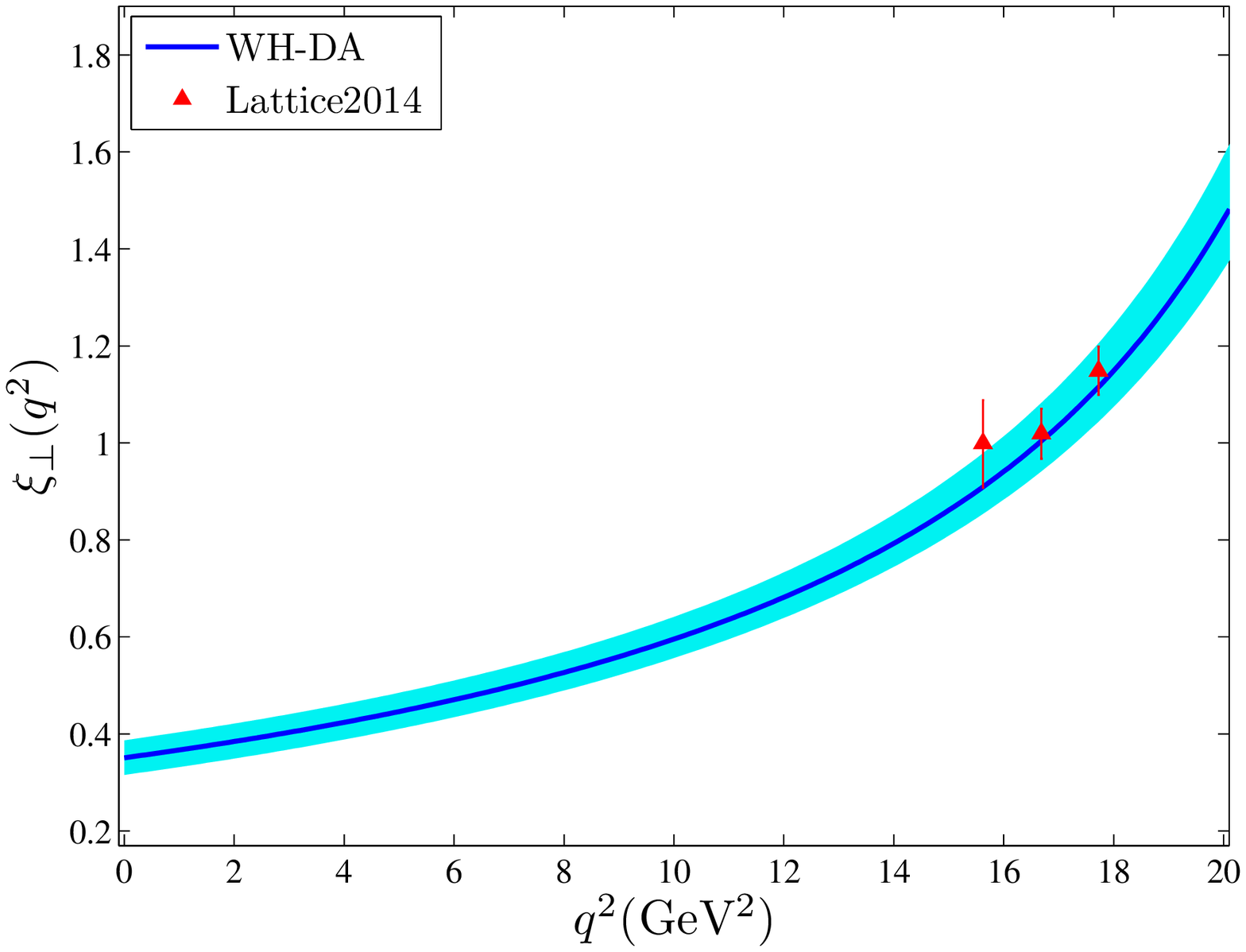}
\caption{The extrapolated $B\to K^*$ TFFs $\xi_\lambda(q^2)$ based on the present LCSR predictions. The shaded bands stand for the theoretical errors. The lattice QCD~\cite{Horgan:2013hoa} prediction has also been presented.}\label{SFFs}
\end{figure}

The LCSR approach is applicable in large and intermediate recoil region, $0\leq q^2 \leq 15~{\rm GeV}^2$. We extrapolate its prediction to the physically allowed $q^2$-region by using a simplified series expansion~\cite{Khodjamirian:2010vf, Bourrely:2008za}, which is based on a rapidly converging series over the parameter $z(t)$, i.e.
\begin{eqnarray}
z(t)=\frac{\sqrt{t_+ - t}-\sqrt{t_+ - t_0}}{\sqrt{t_+ - t}+\sqrt{t_+ - t_0}},
\end{eqnarray}
where $t_\pm=(m_B\pm m_{K^*})^2$ and $t_0=t_+(1-\sqrt{1-t_-/t_+})$. The form factors are them expanded as
\begin{eqnarray}
F_i(q^2)=\frac1{1-q^2/m_{R,i}^2}\sum_{k=0,1,2} a_k^i [z(q^2)-z(0)]^k , \label{extrapolation}
\end{eqnarray}
where $F_i$ stand for the TFFs $\xi_\lambda(q^2)$, and the resonance masses $m_{R,i}$ can be found in Ref.\cite{Straub:2015ica}. The coefficients $a_0^{i}=F_{i}(0)$. The parameters $a_1^i$ and $a_2^i$ are determined by requiring the ``quality of fit ($\Delta$)" to be less than one~\cite{Ball:2004rg}, which is defined as $\Delta={\sum_t\left|F_i(t)-F_i^{\rm fit}(t)\right|}/{\sum_t\left|F_i(t)\right|}\times 100\%$, where $t\in[0,\frac{1}{2},\ldots,\frac{27}{2},14]{\rm GeV}^2$. The extrapolated $B\to K^*$ TFFs are presented in Fig.\ref{SFFs}, in which the lattice QCD prediction~\cite{Horgan:2013hoa} have also been presented. Fig.\ref{SFFs} shows that our present LCSR predictions are consistent with the lattice QCD predictions within errors. In the following, we adopt the extrapolated TFFs to study the forward-backward asymmetry and the isospin asymmetry for the $B\to K^* \mu^+ \mu^-$ decay.

\subsection{The forward-backward and the isospin asymmetries of the $B\to K^* \mu^+ \mu^-$ decay}

\begin{table}[htb]
\centering
\begin{tabular}{ccccccc}
\hline
 & $\bar C_1$ & $\bar C_2$ & $\bar C_3$ & $\bar C_4$ & $\bar C_5$ & $\bar C_6$ \\
$\mu_b$ & -0.148 & 1.060 & 0.012 & -0.035 & 0.010 & -0.039 \\
$\mu_h$ & -0.342 & 1.158 & 0.022 & -0.063 & 0.018 & -0.091 \\
\hline
 & $C_7^{\rm eff}$ & $C_8^{\rm eff}$ & $C_9$ & $C_{10}$\\
$\mu_b$ & -0.307 & -0.169 & 4.238 & -4.641 \\
$\mu_h$ & -0.359 & -0.211 & 4.502 & -4.641\\
\hline
\end{tabular}
\caption{Central values of the Wilson coefficients at the Next-to-Leading Log accuracy at the scale $\mu_b=4.60{\rm GeV}$ and $\mu_h = 1.52{\rm GeV}$, respectively.} \label{Tab:WilsonSM}
\end{table}

The Wilson coefficients are scale-dependent, whose values at the lower scales such as the typical momentum flow of the $B$-meson decay, $\mu_b \sim m_b$, or the hadronic scale, $\mu_h=\sqrt{\Lambda_h \mu_b}$~\cite{Ahmady:2014cpa},  can be derived from the values at the weak scale $\mu_W = {\cal O}(M_W)$ via renormalization group equation. At the electroweak scale, the Wilson coefficients can be written as
\begin{eqnarray}
&&C_i(\mu_W) = C_i(\mu_W)_{\rm SM} + \delta C_i(\mu_W)_{\rm H} + \delta C_i(\mu_W)_{\rm SUSY}
\nonumber \\[0.15em]
&&\qquad= C_i^{(0)}(\mu_W)_{\rm SM}+ \delta C_i^{(0)}(\mu_W)_{\rm H}+\delta C_i^{(0)}(\mu_W)_{\rm SUSY}
\nonumber \\
&&\qquad+ {\alpha_s(\mu_W) \over 4 \pi} \,
\bigg[C_i^{(1)} (\mu_W)_{\rm SM}+\delta C_i^{(1)}(\mu_W)_{\rm H} \nonumber \\
&&\qquad+
\delta C_i^{(1)}(\mu_W)_{\rm SUSY} \bigg] ,
\label{deltadef}
\end{eqnarray}
where $i=(1,\cdots,10)$. The expression of $C_i^{(0)}$ and $C_i^{(1)}$ can be found in Ref.\cite{Buras:1998raa} and those of $\delta C_{7,8}^{(0,1)}$ can be found in Ref.\cite{Ciuchini:1998xy}. Up to NLO level, the first six Wilson coefficients $\bar C_i (\mu)$ can be rewritten as
\begin{eqnarray}
\bar C_i (\mu) = C_i (\mu) + \frac{\alpha_s(\mu)}{4\pi} T_{ij} C_j(\mu) \quad i\in \{1,2,\ldots,6\},
\end{eqnarray}
in which $T_{ij}$  is the transformation matrix~\cite{Beneke:2001at}. The SUSY contributions to $C_9$ and $C_{10}$ have been calculated in Refs.\cite{Cho:1996we, Hewett:1996ct}. The interactions among the charged Higgs and the up-type quarks, which are from the SUSY model or the SM models with two Higgs doublet, may have sizable contributions, those terms are represented by a subscript ${\rm H}$. Because the NLO SUSY contribution to the four-quark penguin operators are also sizable, we treat it as $\delta C_i^{(1)}(\mu_W)_{\rm H}$~\cite{Bobeth:1999ww}. The central values of NLL Wilson coefficients at scale $\mu_b=4.6{\rm GeV}$ and $\mu_h = 1.52{\rm GeV}$ are presented in Table~\ref{Tab:WilsonSM}, where as suggested by Ref.\cite{Beneke:2001at}, we take $C_{7}^{\rm eff}=C_7 -(4\bar{C}_3-\bar{C}_5)/9 -(4\bar{C}-\bar{C}_6)/3$ and $C_{8}^{\rm eff}=C_8+(4\bar{C}_{3}-\bar{C}_5)/3$.

We adopt MSSM with MFV as a typical SUSY model to probe the possible new physics effect. The basic input for the SUSY parameters is the ratio of the vacuum expectation values of the Higgs doublet, i.e. $\tan\beta$, and we take $\tan\beta\in[2,40]$ to do the discussion. A larger $\tan\beta$ could lead to a flip of sign for $C_{7,8}^{\rm eff}$~\cite{Degrassi:2000qf}, which arouses people's great interests. The behaviours of the two ranges as a function of the free parameters are quite different, we call the model for $\tan\beta\in[2,10]$ as MFV-I and the model for $\tan\beta=40$ as MFV-II. The mechanisms that enhance the SUSY contribution to $C_7^{\rm eff}$ at large $\tan\beta$ are not working for $C_{9,10}$~\cite{Cho:1996we, Hewett:1996ct}. For instance, the charged Higgs contribution dominant for $C_7^{\rm eff}$ at large $\tan\beta$ is suppressed for $C_{9,10}$, and the modifications for the forward-backward and the isospin asymmetries shall be mainly due to the new physics contributions to $C_{7,8}^{\rm eff}$. In allowable parameter space, the ranges of the new physics part of $C_{7,8}^{\rm eff}$ at the scales $\mu_b$ and $\mu_h$ are
\begin{eqnarray}
&&\delta C_7^{\rm I}(\mu_b)\in[-0.028,0.168],\,
      \delta C_7^{\rm I}(\mu_h)\in[-0.033,0.196],\,  \nonumber\\
&&\delta C_8^{\rm I}(\mu_b)\in[-0.256,0.043],\,
      \delta C_8^{\rm I}(\mu_h)\in[-0.320,0.053],\, \nonumber\\
&& \delta C_7^{\rm II}(\mu_h)\in[0.963,2.023],\,
       \delta C_7^{\rm II}(\mu_b)\in[0.825,1.733],\nonumber \\
&& \delta C_8^{\rm II}(\mu_b)\in[0.284,0.810],
       \delta C_8^{\rm II}(\mu_h)\in[0.355,1.011]. \nonumber
\end{eqnarray}
where I corresponds to the MFV-I and II corresponds to the MFV-II, respectively.

\begin{figure*}[htb]
\begin{center}
\includegraphics[width=0.3\textwidth]{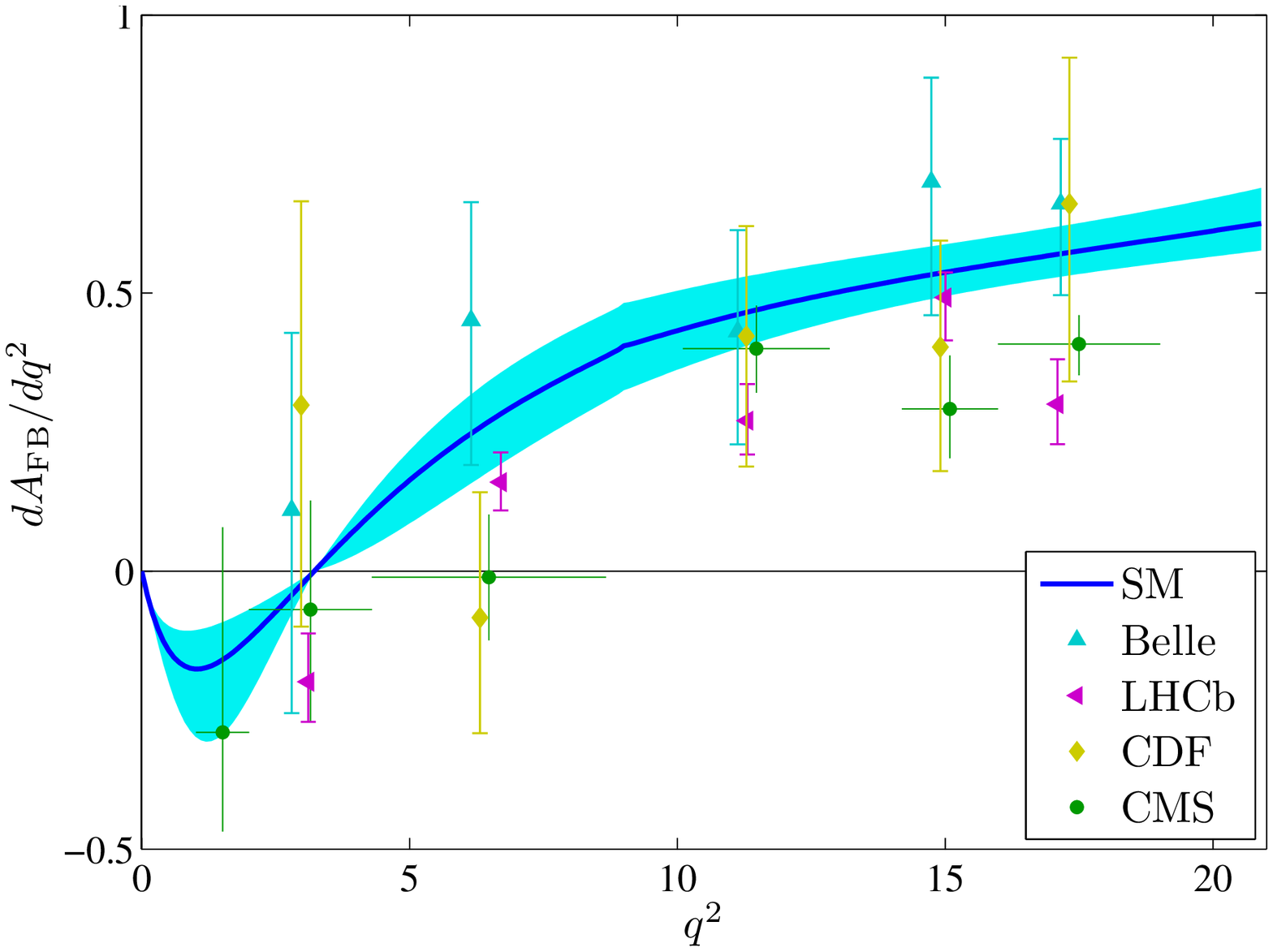}
\includegraphics[width=0.3\textwidth]{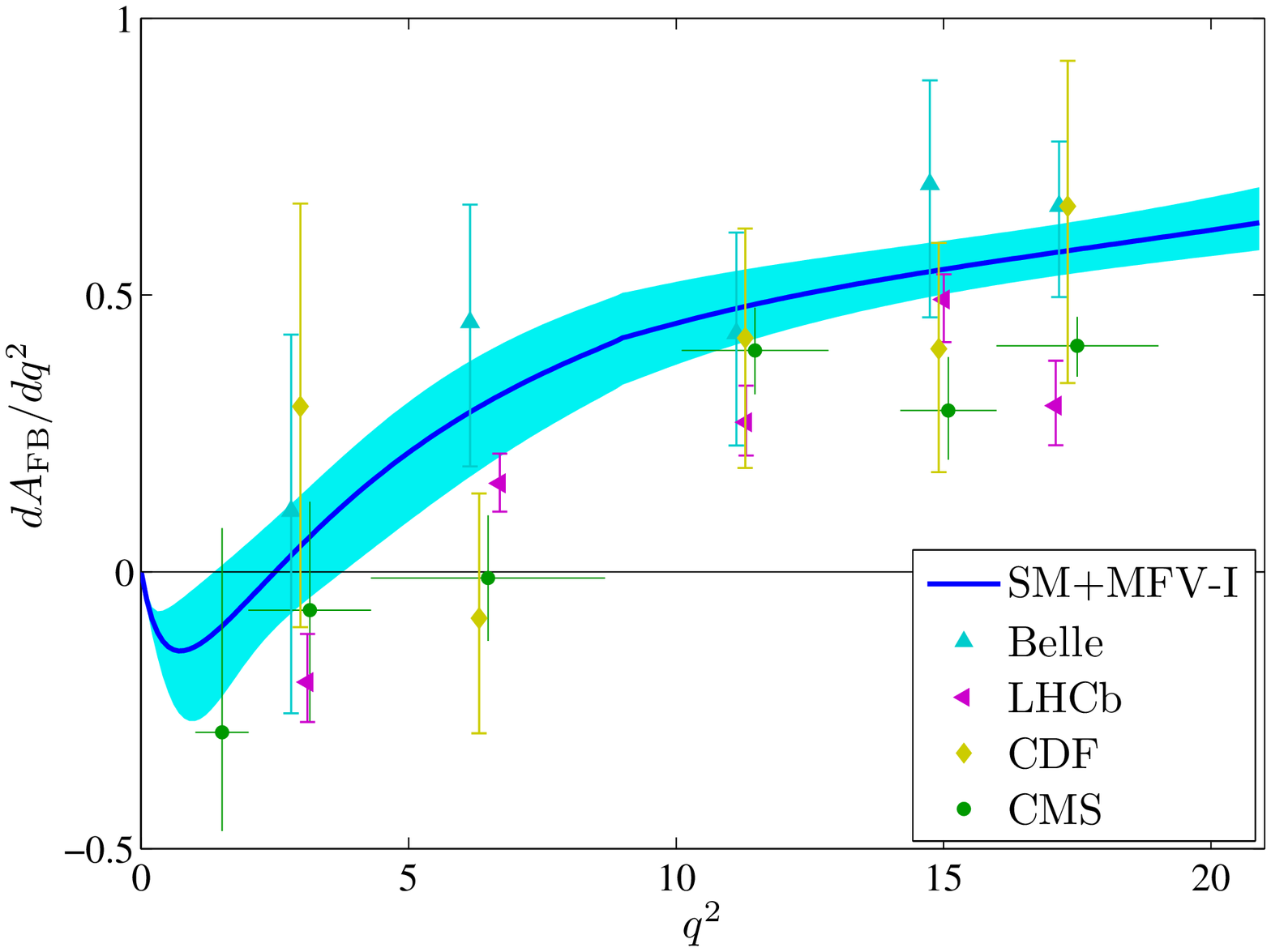}
\includegraphics[width=0.3\textwidth]{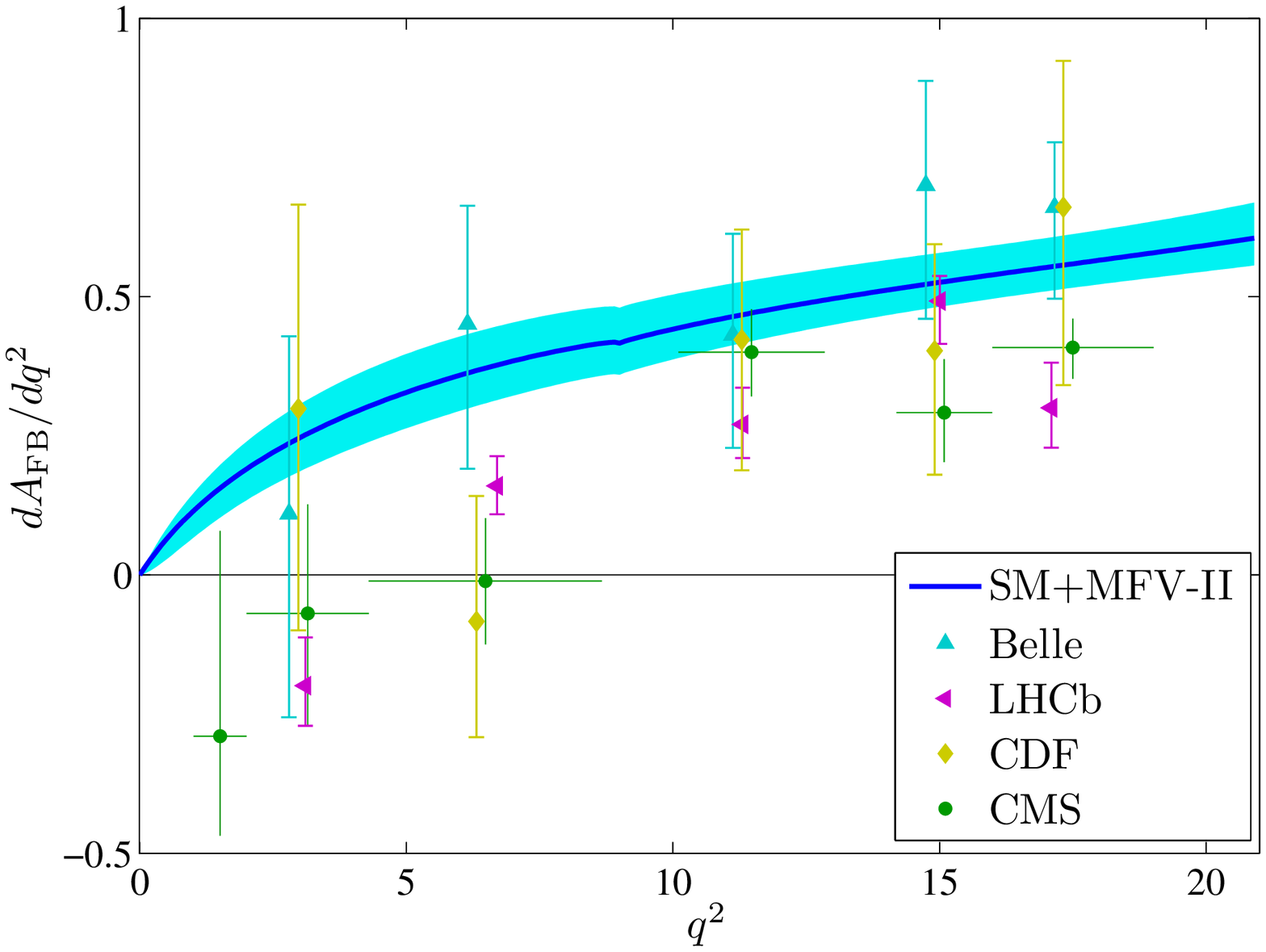}
\caption{The forward-backward asymmetries, where the shaded bands stand for the theoretical errors. The measurements from the Belle~\cite{Wei:2009zv}, the CDF~\cite{Aaltonen:2011ja}, the LHCb~\cite{Aaij:2014pli}, and the CMS~\cite{Chatrchyan:2013cda} collaborations have also been presented.}
\label{AFB1}
\end{center}
\end{figure*}

By using the parameters in the SM and the MSSM MFV scenarios allowed from the constraints discussed in above, we give our prediction on the forward-backward and the isospin asymmetries of the $B\to K^* \mu^+ \mu^-$ decay in the following paragraphs.

The differential distributions for the forward-backward asymmetry are presented in Fig.\ref{AFB1}, where the measurements from the Belle~\cite{Wei:2009zv}, the CDF~\cite{Aaltonen:2011ja}, the LHCb~\cite{Aaij:2014pli} and the CMS~\cite{Chatrchyan:2013cda} collaborations have been presented. Fig.\ref{AFB1} shows that
\begin{itemize}
\item In large $q^2$-region, $q^2\in[6,19]~{\rm GeV^2}$, three curves behave closely, all of which increase monotonously with the increment of $q^2$. The magnitude of the MSSM MFV-I terms generally decreases with the increment of $q^2$, i.e. its portion to the corresponding SM error becomes $34\%$, $11\%$, $4\%$, and $\le 1\%$ for $q^2=6~{\rm GeV^2}$, $8~{\rm GeV^2}$, $10~{\rm GeV^2}$, and $20~{\rm GeV^2}$, respectively.  The magnitude of the MSSM MFV-II terms shall first increase and then decreases with the increment of $q^2$, i.e. its portion to the corresponding SM error becomes $2\%$, $14\%$, $13\%$, $5\%$, and $\le 1\%$ for $q^2=6~{\rm GeV^2}$, $8~{\rm GeV^2}$, $10~{\rm GeV^2}$, $12~{\rm GeV^2}$, and $20~{\rm GeV^2}$, respectively.

    Such a smaller effect to the SM prediction at the large-$q^2$ region indicates that one can not distinguish those MSSM MFV models with the SM one in the large $q^2$-region. In large $q^2$-region, the predicted forward-backward asymmetry agrees with the Belle~\cite{Wei:2009zv} and the CDF~\cite{Aaltonen:2011ja} measurements within errors. However even by including the MSSM MFV-I or MFV-II terms, we still cannot explain the trends of a smaller forward-backward asymmetry around $q^2>16~{\rm GeV^2}$ as indicated by the LHCb~\cite{Aaij:2014pli} and the CMS~\cite{Chatrchyan:2013cda} measurements. Thus we need new SUSY models to explain this discrepancy, or we need more data to confirm those measurements.

\item Main differences among various models lie in low $q^2$-region, e.g. $q^2 \leq 6~{\rm GeV}^2$, indicating the MSSM effects could be important and sizable. The SM prediction has a cross-over around $q^2\sim 3.2~{\rm GeV}^2$, which shifts to a smaller value for MSSM MFV-I. The forward-backward asymmetries of SM and MSSM MFV-I behave closely in shape, both of which are negative for small $q^2$-region and are consistent with the measurements. Meanwhile, the forward-backward asymmetries of MSSM MFV-II are always positive in low $q^2$-region, which is due to the flip of sign for $C_{7,8}^{\rm eff}$ at large $\tan\beta$ and is out of the LHCb and CMS measurements. Thus the present data prefers a smaller $\tan\beta$, i.e. MSSM MFV-I. Due to different behaviors of the forward-backward asymmetries under MSSM MFV-I and MFV-II, the more precise measurements in low $q^2$-region shall be helpful for constraining a more reliable range for the key MSSM parameter $\tan\beta$.
\end{itemize}

\begin{table}[tb]
\centering
\begin{tabular}{cc}
\hline
                  ~~~ ~~~          & ~~~$A_{\rm FB}(q^2\in[1,6]{\rm GeV}^2)$~~~          \\
\hline
SM         & $0.121^{+0.212}_{-0.357}$\\
SM+MFV-I      & $0.425^{+0.225 + 0.395}_{-0.369-0.424}$\\
SM+MFV-II     & $1.297^{+0.203 + 0.227}_{-0.189-0.213}$   \\
Belle~\cite{Wei:2009zv}      & $0.26^{+0.27}_{-0.30}\pm 0.07$ \\
CMS~\cite{Chatrchyan:2013cda}& $-0.05\pm 0.03$ \\
ATLAS~\cite{ATLAS:2013ola}     & $0.07\pm0.20\pm0.07$   \\
\hline
\end{tabular}
\caption{Integrated forward-backward asymmetries for $q^2\in[1,6]{\rm GeV}^2$ under the SM, the MSSM MFV-I and the MFV-II, respectively. The results for the, the Belle~\cite{Wei:2009zv}, the CMS~\cite{Chatrchyan:2013cda}, and ATLAS~\cite{ATLAS:2013ola} measurements are also presented.}
\label{Tab:AFB}
\end{table}

Next, we present the integrated forward-backward asymmetry for $q^2\in[1,6]{\rm GeV}^2$ in Table~\ref{Tab:AFB}. Here the first uncertainty is the SM error which is mainly from the LCSR predictions and the second one is the MSSM MFV-I or MFV-II error which is dominated by the possible choices of $C_{7,8}^{\rm eff}$. In Table~\ref{Tab:AFB} We also present the Belle~\cite{Wei:2009zv}, the CMS~\cite{Chatrchyan:2013cda}, and the ATLAS~\cite{ATLAS:2013ola} data as a comparison.  Table~\ref{Tab:AFB} confirms our above observation that the MSSM MFV-I gives SM-like prediction, both of which are consistent with the measurements within errors; while, the MSSM MFV-II prefers a quite large asymmetry $A_{\rm FB}$.

\begin{figure*}[htb]
\begin{center}
\includegraphics[width=0.3\textwidth]{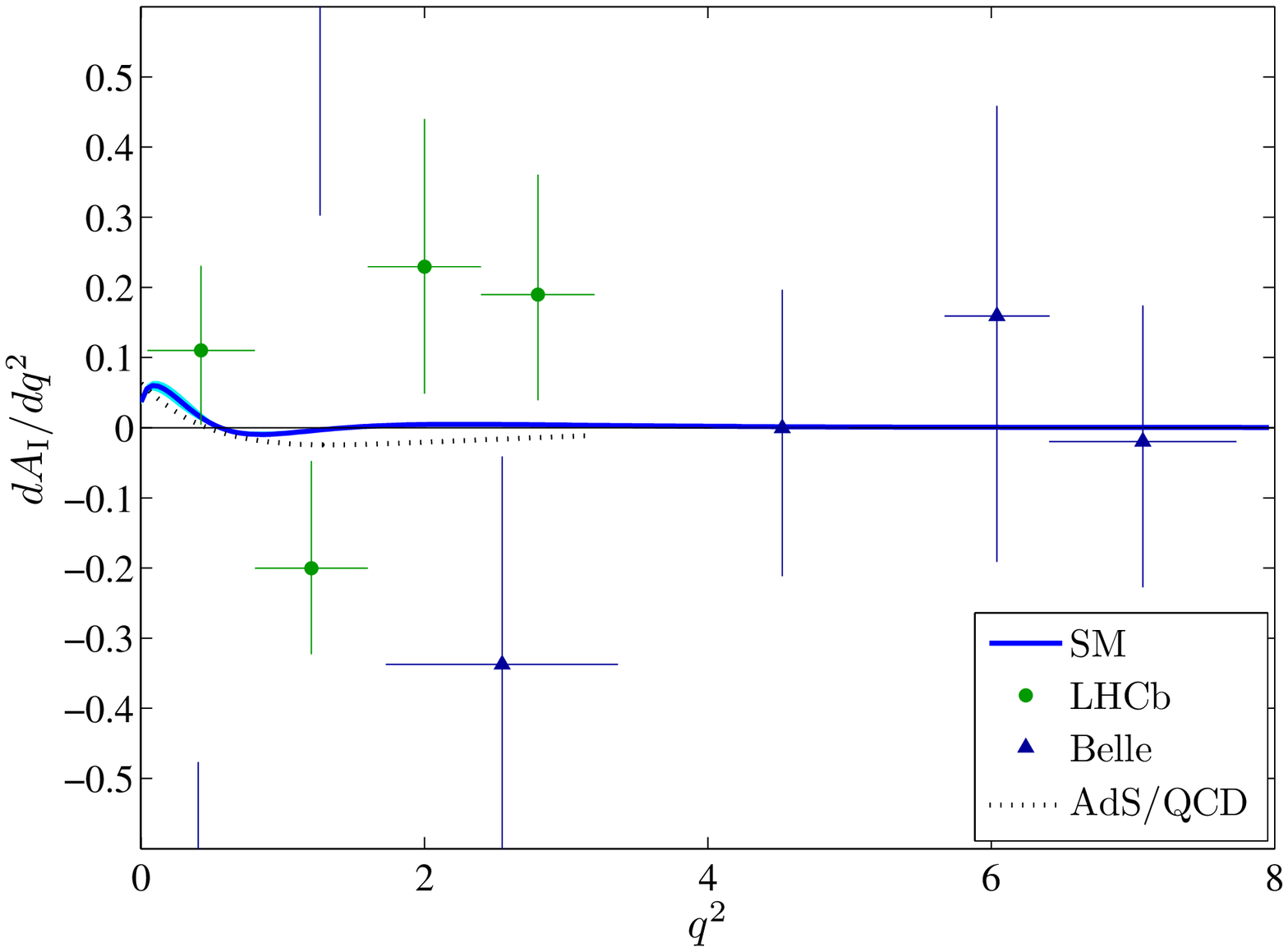}
\includegraphics[width=0.3\textwidth]{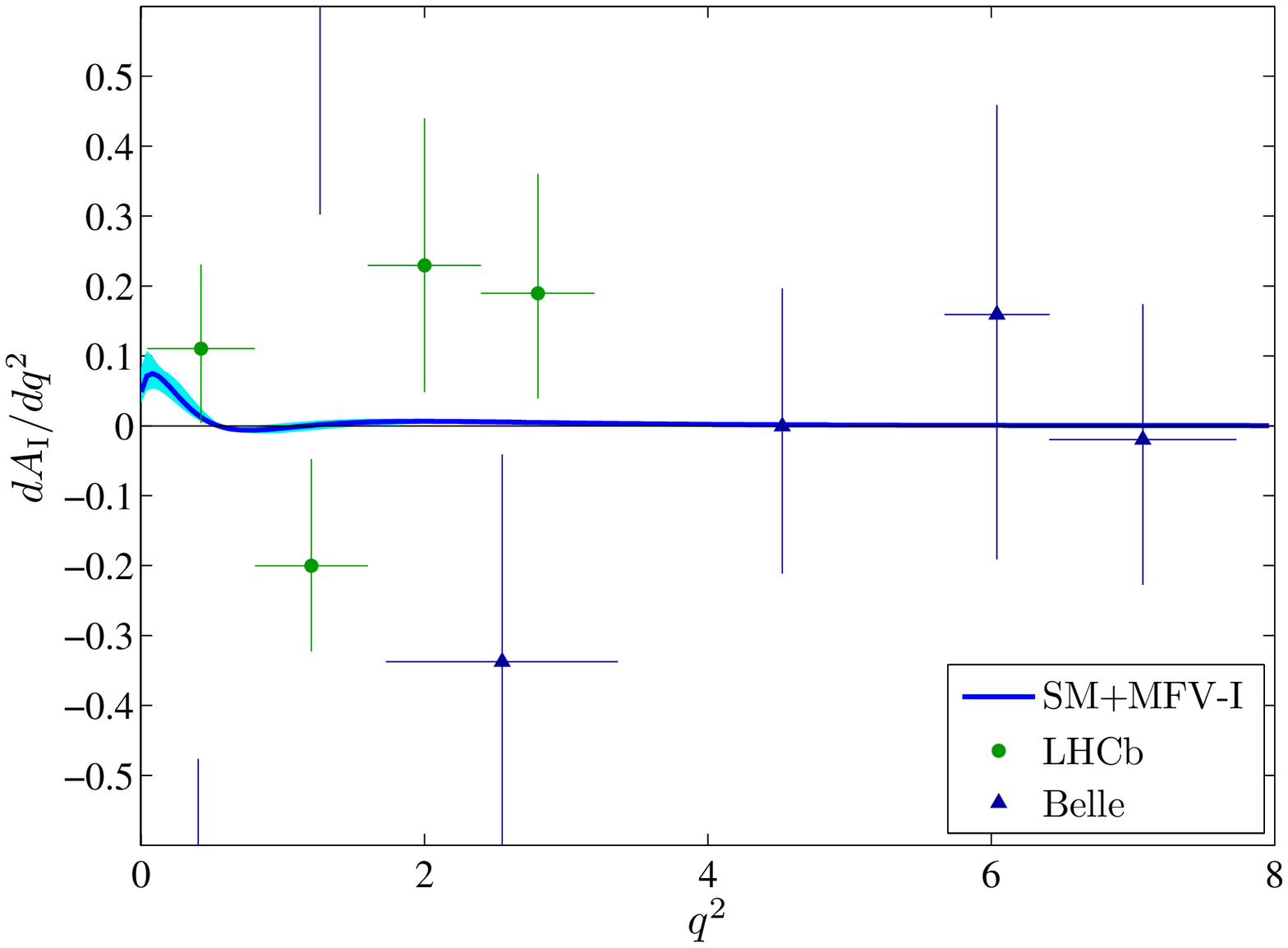}
\includegraphics[width=0.3\textwidth]{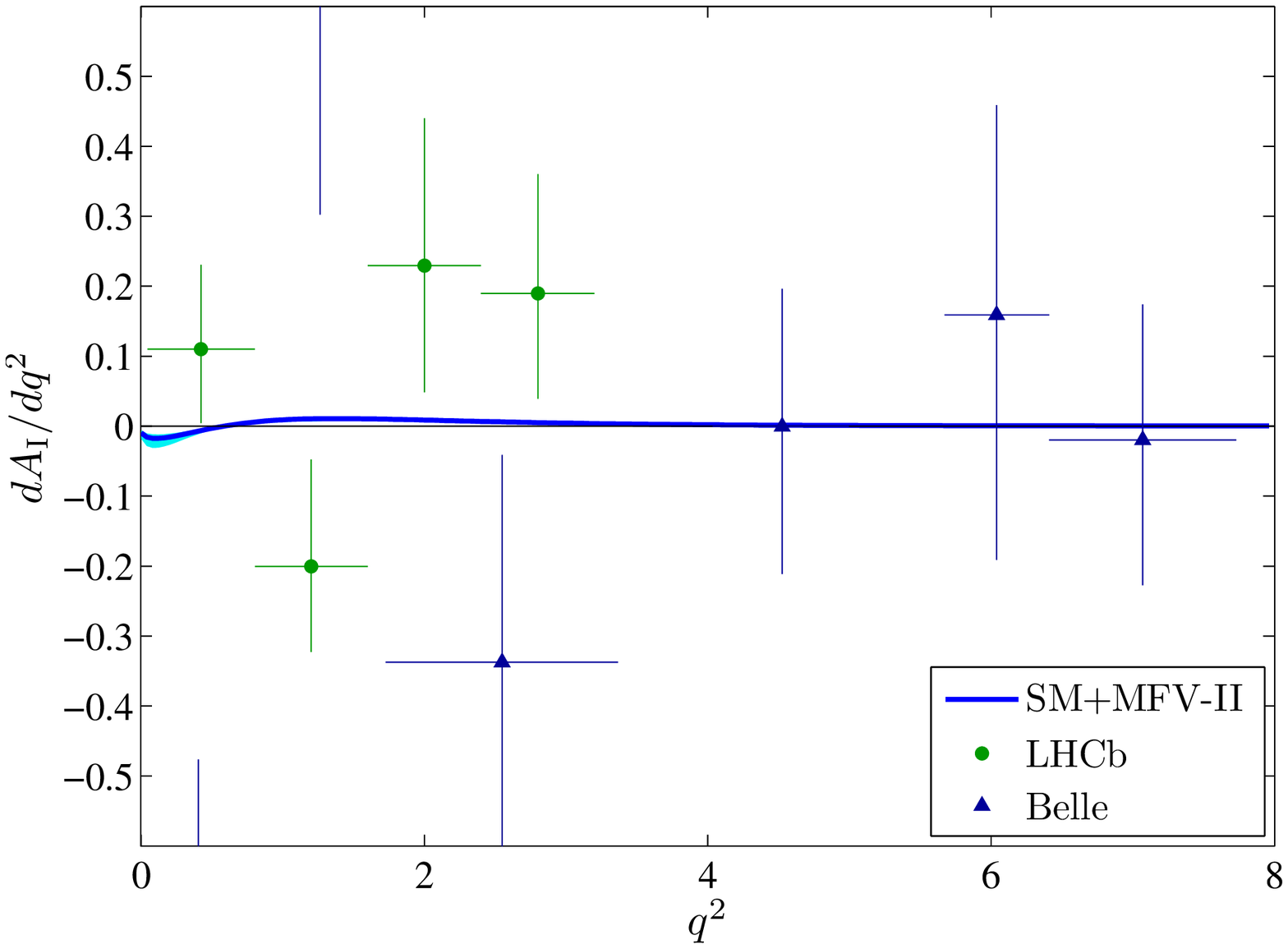}
\caption{The isospin asymmetry of the $B\to K^* \mu^+ \mu^-$ decay. The SM prediction by using the AdS/QCD LCDA~\cite{Ahmady:2014cpa} is presented in the first diagram as a comparison. The measurements from the Belle~\cite{Wei:2009zv} and the LHCb~\cite{Aaij:2014pli} collaborations are also presented.}
\label{Fig:AI1}
\end{center}
\end{figure*}

Finally, we present the differential distribution for the isospin asymmetry of the $B\to K^* \mu^+ \mu^-$ decay in Fig.\ref{Fig:AI1}. The first sub-figure of Fig.\ref{Fig:AI1} is the SM prediction, where the uncertainties are squared averages of all SM uncertainties which are dominated by the LCSR prediction of the TFFs. The SM prediction by using the AdS/QCD-LCDA for $q^2\leq 8~{\rm GeV^2}$~\cite{Ahmady:2014cpa} is presented as a comparison. It shows that even though the AdS/QCD-LCDA is quite different from our present choice of LCDA, leading to different TFFs, the isospin asymmetries behave closely, which are about $0.5-0.6$ for $q^2\to 0$ and tend to zero for larger $q^2$-region. The second and third sub-figures are for the SM + MSSM MFV-I and SM + MSSM MFV-II scenarios, in which the uncertainties are combined ones of SM and MSSM input parameters. The Belle~\cite{Wei:2009zv} and the LHCb~\cite{Aaij:2014pli} measurements are included in those figures.

The narrow error bands for the isospin asymmetries under the SM, the MSSM MFV-I and the MSSM MFV-II are due to the fact that the isospin asymmetry is dominated by the penguin coefficients $C_{3-6}$ which are only slightly affected by both the SM and MSSM MFV input parameters. Because the measurements are still of large errors, all the predictions are consistent with data. More subtly, the SM and MSSM MFV-I isospins are positive for $q^2 \in[0, 1.4]~{\rm GeV}^2$ and negative for $q^2 \in[1.5, 2.9]~{\rm GeV}^2$, agreeing with the LHCb trends; The MSSM MFV-II also leads to a smaller flip of sign of the isospin asymmetry which is negative at $q^2 \in[0, 1.4]~{\rm GeV}^2$.

\subsection{Branching ratio and longitudinal polarization fraction for $B\to K^* \nu\bar\nu$}

The $B\to K^* \nu\bar\nu$ decay has the virtue that the angular distribution of the $K^*$ decay products allows to extract information on the $K^*$ polarization, similar to the $B\to K^* \mu^+\mu^-$ decays. The longitudinal and transverse differential distributions versus $q^2$, the square of the invariant mass of the $\nu\bar\nu$ pair, is given as~\cite{Colangelo:1996ay, Melikhov:1998ug, Buchalla:2000sk}
\begin{eqnarray}
\frac{d\Gamma_L}{dq^2} &=& N |H_0(q^2)|^2,\nonumber\\
\frac{d\Gamma_T}{dq^2} &=& N [| H_\bot(q^2)|^2  + \vert H_\|(q^2) |^2 ]  \label{dGamma}
\end{eqnarray}
with the coefficient $N= \frac{G_F^2 |V_{tb} V_{ts}^*|^2 \alpha_{\rm em}^2}{256 \pi^5 m_B} \lambda^{1/2} (q^2) q^2$. The hadronic transversity amplitudes $H_{\bot,\|,0}(q^2)$ are
\begin{eqnarray}
H_\perp(q^2) &=& \frac{\sqrt{2}(C_L+C_R)m_B^2\lambda^{1/2}(q^2)} {m_B+m_{K^\ast}}V(q^2),\nonumber\\
H_\parallel(q^2) &=& \sqrt{2}(C_L-C_R)(m_B+m_{K^\ast})A_1(q^2),\nonumber\\
H_0(q^2) &=& -\frac{1}{2m_{K^\ast}\sqrt{q^2}}(C_L-C_R)\bigg[(m_B+m_{K^\ast})\nonumber\\
&&\!\!\!\!\!\!\!\!\!\!\!\!\!\!\!\!\! (m_B^2 -m_{K^*}^2 - q^2) A_1 (q^2) - \frac{m_B^4\lambda(q^2)} {m_B + m_{K^*}} A_2(q^2)\bigg].
\end{eqnarray}
The total differential decay width $d\Gamma/q^2 = d\Gamma_L/q^2 + d\Gamma_T/q^2$. To calculate the branching ratios, we use the average value from the $B^\pm$ lifetime $\tau_{B^+}$ and the $B^0$ lifetime $\tau_{B^0}$ for $B\to K^* \nu\bar\nu$ decay. Meanwhile, the $K^*$-meson longitudinal and transverse polarization fraction $F_{L, T}$ are defined as
\begin{eqnarray}
F_{L,T} = \frac{d\Gamma_{L,T}/dq^2}{d\Gamma/dq^2}, \label{FL}
\end{eqnarray}
which satisfy $F_L + F_T =1$. The TFFs $A_{1,2}(q^2)$ and $V(q^2)$ have also been calculated by using a right-handed chiral correlator under the LCSR approach~\cite{Fu:2014uea, Cheng:2017bzz}.

Principally, the Wilson coefficients $C_L$ and $C_R$ are complex. One usually defines two real parameters,
\begin{eqnarray}
\epsilon = \frac{\sqrt{|C_L|^2+|C_R|^2}}{|C_L^{\rm SM}|^2}, \eta = - \frac{{\rm Re}(C_L C_R^*)}{|C_L|^2+|C_R|^2},
\end{eqnarray}
and the differential decay branching ratio and longitudinal polarization fraction can be expressed as
\begin{eqnarray}
\frac{d{\cal B}(B\to K^*\nu\bar\nu)}{dq^2} &= &\frac{d{\cal B}^{\rm SM}(B\to K^*\nu\bar\nu)}{dq^2}(1+1.31\eta)\epsilon^2 \nonumber\\
F_L(B\to K^*\nu\bar\nu) &= & F_L^{\rm SM}(B\to K^*\nu\bar\nu) \frac{1+2\eta}{1+1.31\eta}
\end{eqnarray}
The Wilson coefficient $C^{\rm SM}_R$ for the SM is negligibly small, leading to $\eta^{\rm SM}\simeq0$. The Wilson coefficient $C^{\rm SM}_L$ for the SM has been calculated at the next-to-leading order QCD corrections~\cite{Buchalla:1998ba, Misiak:1999yg}, which gives $C^{\rm SM}_L=-X(x_t)/\sin^2\theta_W$, where $x_t=m_t^2/m_W^2$ and $X(x_i)$ is the corresponding loop function which gives $C_L^{\rm SM} = -6.38(6)$~\cite{Altmannshofer:2009ma}.

Different to the above considered case of two leptons in final state which uses MSSM with MFV to deal with the new physics effect, as suggested by Ref.\cite{Altmannshofer:2009ma}, we adopt the MSSM with generic flavour violating (GFV) to deal with the two Wilson coefficients $C_L$ and $C_R$ for the present case of two neutrinos in final state. In this model, the MSSM contributions to $C_R$ turn out to be very small, which implies that $\eta \simeq 0$, thus leads to a SM-like prediction on $F_L(q^2)$, i.e. $F_L(q^2)\simeq F^{\rm SM}_L(q^2)$. Thus one cannot use the observable $F_L(q^2)$ along to probe the MSSM.

\begin{figure}[htb]
\begin{center}
\includegraphics[width=0.45\textwidth]{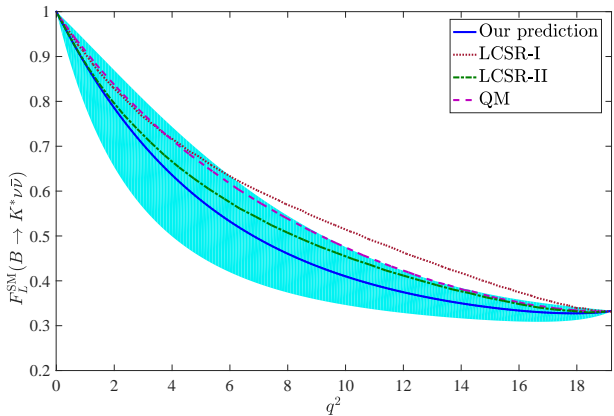}
\caption{The $K^*$-meson longitudinal polarization fraction $F^{\rm SM}_{L}$ of the $B\to K^*\nu\bar\nu$ decay within the SM. The solid line together with the shaded band is our present prediction with the uncertainties from the TFFs $A_{1,2}$ and $V$. The dashed, the dash-dot and the dotted liens are for the LCSR-I result~\cite{Altmannshofer:2008dz}, LCSR-II result~\cite{Ball:2004rg} and the QM result~\cite{Melikhov:1996du}, respectively. }
\label{Fig:FL}
\end{center}
\end{figure}

\begin{figure}[htb]
\begin{center}
\includegraphics[width=0.45\textwidth]{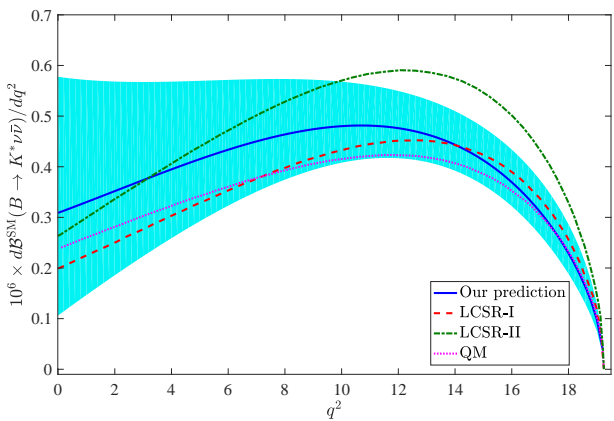}
\caption{A comparison of the SM differential branching ratio of the $B\to K^*\nu\bar\nu$ decay under various approaches, in which the shaded band stands for the uncertainties of our prediction from the TFFs $A_{1,2}$ and $V$ and the Wilson coefficient $C_L$.  }
\label{Fig:dBnew2}
\end{center}
\end{figure}

We present our prediction of the $K^*$-meson longitudinal polarization fraction $F^{\rm SM}_{L}$ in Fig.\ref{Fig:FL}. As a comparison we also present the QM result~\cite{Melikhov:1996du} and the other two LCSR predictions, i.e. LCSR-I~\cite{Altmannshofer:2008dz}and LCSR-II~\cite{Ball:2004rg} in the figure. Fig.\ref{Fig:FL} shows our results are consistent with LCSR-II and QM results within reasonable errors in whole $q^2$-region, while the LCSR-I has a larger $F_L(q^2)$ in intermediate and large $q^2$-region such as $6\;{\rm GeV}^2 < q^2 < (m_B-m_K^*)^2$. We present a comparison of the SM differential branching ratio of the $B\to K^*\nu\bar\nu$ decay under various approaches in Fig.\ref{Fig:dBnew2}. It shows that different TFFs leads to different behaviors, the LCSR-I and QM results agree with our differential branching ratio within errors; while the LCSR-II agrees with our prediction only for small $q^2$-region, e.g. $0<q^2<10{\rm GeV}^2$. Thus a more accurate prediction on the $B\to K^*$ TFF shall be helpful for a more accurate SM prediction.

The possible visible MSSM effects in $C_L$ are generated by chargino contributions through a large $(\delta_u^{RL})_{32}$ mass insertion. Because those chargino contributions are not sensitive to the choice of $\tan \beta$, we choose to work in the low $\tan\beta$ regime, i.e. $\tan\beta = 5$. The necessary inputs for the MSSM with GFV can be found in the Ref.\cite{Altmannshofer:2009ma}, in which there are two typical sets of parameters, and we call them as MSSM GFV-I and GFV-II, respectively.

\begin{figure}[htb]
\begin{center}
\includegraphics[width=0.45\textwidth]{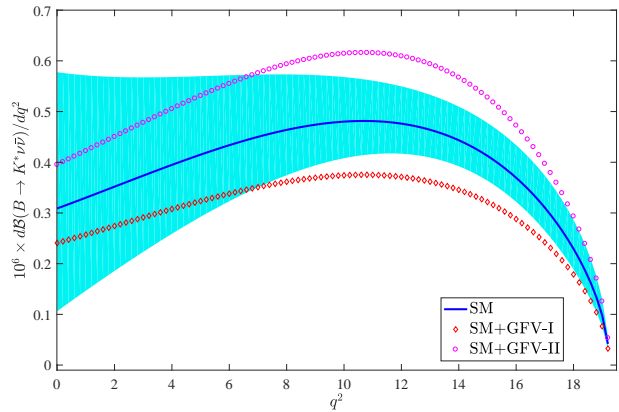}
\caption{The differential branching ratio of the $B\to K^*\nu\bar\nu$ decay under various models, in which the shaded band stands for the uncertainties from the TFFs $A_{1,2}$ and $V$ and the Wilson coefficient $C_L$. }
\label{Fig:dBnew}
\end{center}
\end{figure}

By using the parameters in the SM and the MSSM with GFV scenarios allowed from the constraints discussed in above, we give our prediction on the differential branching ratio in Fig.\ref{Fig:dBnew}. The SM plus MSSM GFV-I and GFV-II predictions are consistent with the SM prediction in low $q^2$-region, but are different at high $q^2$-region, e.g. $q^2>7\;{\rm GeV}^2$. Thus a careful measurement at high $q^2$-region could be helpful to clarify whether there is new physics, and which one, MSSM GFV-I or MSSM GFV-II, is more preferable.

\begin{table}[htb]
\centering
\begin{tabular}{ccc}
\hline
     ~~~ & ~~~${\cal B}\times 10^6$~~~ &~~~ $\langle F_L\rangle$ ~~~\\ \hline
SM & $7.60^{+2.16}_{-1.70}$ & $0.49^{+0.09}_{-0.10}$  \\
SM+GFV-I & $5.92^{+1.68}_{-1.33}$ & -- \\
SM+GFV-II & $9.72^{+2.76}_{-2.18}$& --\\
Belle~\cite{Grygier:2017tzo} & $ <18 $ & -- \\
ABSW~\cite{Altmannshofer:2009ma} (SM)           & $6.8^{+1.0}_{-1.1}$  & $0.54(1)$           \\
ABSW~\cite{Altmannshofer:2009ma} (GFV-I)       & $5.3$           & --           \\
ABSW~\cite{Altmannshofer:2009ma} (GFV-II)      & $8.7$           & --           \\
NWA(SM)~\cite{Das:2017ebx}  & $9.49(101)$ & 0.49(4) \\
\hline
\end{tabular}
\caption{The branching ratio $\cal B$ (in unit $10^{-6}$) and the longitudinal polarization fraction $\langle F_L \rangle$ of $B\to K^*\nu\bar\nu$. The errors are squared average of all mentioned error sources. The predictions of the Belle Collaboration~\cite{Grygier:2017tzo}, the ABSW~\cite{Altmannshofer:2009ma} and NWA~\cite{Das:2017ebx} are presented as a comparison. } \label{Tab:dBFL}
\end{table}

To have a clear look at the differences among different models and approaches, we integrate the momentum transfer in whole physical region $0<q^2<(m_B-m_K^*)^2$ to get the total branching ratio and the $q^2$-integrated form of $F_L$, which is define as
\begin{equation}
\langle F_L \rangle = \frac{\Gamma_L}{\Gamma},
\end{equation}
where
\begin{equation}
\Gamma_{(L)} = \int_0^{(m_B-m_K^*)^2} dq^2 \frac{d\Gamma_{(L)}}{dq^2}.
\end{equation}
We present the results for the branching ratio $\cal B$ and the longitudinal polarization fraction $\langle F_L \rangle$ of $B\to K^*\nu\bar\nu$ in Table~\ref{Tab:dBFL}. As a comparison, we also present the 2017 Belle Collaboration measurements~\cite{Grygier:2017tzo}, the results of Ref.\cite{Altmannshofer:2009ma} (ABSW) and the SM prediction of Ref.\cite{Das:2017ebx} (NWA) in Table~\ref{Tab:dBFL}. As mentioned above, the MSSM effect to $F_L$ is negligibly small due to $\eta\to 0$, and the predicted values of $\langle F_L \rangle$ within the SM are listed in the second series at Table~\ref{Tab:dBFL}, which shows that our prediction is close to NWA one. Table~\ref{Tab:dBFL} shows by including the new physics effect, the branching ratio shall be suppressed by $\sim22\%$ and increased by $\sim28\%$ for MSSM GFV-I and GFV-II, respectively. Our SM prediction of the branching ratio $\cal B$ are consistent with the ABSW SM and NWA predictions within errors, all of which agree with the newest upper limit predicted by Belle Collaboration in 2017 $({\cal B}^{Belle}<18\times 10^{-6})$. Thus we still  need more accurate data to draw definite conclusions.

\section{Summary}\label{Sum}

In the paper, we recalculate the $B\to K^*$ TFFs $\xi_{\perp,\|}(q^2)$ by using the LCSR approach, in which a chiral correlator has been adopted to suppress the large uncertainties from the twist-2 and twist-3 structures at the $\delta^1$-order. For each LCSR, except the dominate twist-2 contribution which are proportional to $\phi_{2;K^*}^{\perp}$, the remaining non-zero twist-3 and twist-4 terms as shown by Eqs.(\ref{xi_bot}, \ref{xi_parallel}) shall be at least $\delta^2$-suppressed, which totally only provide less than $10\%$ contributions to the LCSRs. Thus the resultant LCSRs are more accurate than the previous ones derived in the literature. The extrapolated $B\to K^*$ TFFs as shown in Fig.\ref{SFFs} are consistent with the Lattice QCD predictions within errors.  This new achievement helps for probing new physics beyond the SM.

Based on the definitions of the forward-backward and the isospin asymmetries, we calculate their differential distributions over $q^2$ under three models and present our results in Figs.\ref{AFB1} and \ref{Fig:AI1}. The SM and the SM+MSSM MFV-I predictions are consistent with each other; while the SM+MSSM MFV-II prediction shows quite different behavior, especially in low $q^2$-region. Thus a careful comparing with data could be helpful for judging whether we need new physics scenario for those observables or which new physics scenario is more credible:
\begin{itemize}
\item For the forward-backward asymmetry $A_{\rm FB}$, the MSSM MFV-I only slightly changes the SM prediction and does not change its arising trends, both of which agree with the Belle, the CDF and the CMS measurements in low $q^2$-region. On the contrary, due to the flip of sign for $C^{\rm eff}_{7,8}$, the MSSM MFV-II give large corrections to the SM prediction in low $q^2$-region, leading to a positive $A_{\rm FB}$ in whole $q^2$-region. This differences make it possible to draw the conclusion of whether MFV-I or MFV-II is preferable by using more accurate data measured in low $q^2$-region. Table~\ref{Tab:AFB} prefers a small $\tan\beta$ for the MSSM MFV model.

\item For the forward-backward asymmetry $A_{\rm FB}$ at the large $q^2$-region, we have found that the new physics effect shall be suppressed by $1/q^4$ to compare with the SM prediction. Fig.\ref{AFB1} and Fig.\ref{Fig:AI1} show that even by including the MSSM MFV-I or MFV-II terms, we still cannot explain the trends of the smaller forward-backward asymmetry around $q^2>16~{\rm GeV^2}$ as indicated by the present LHCb~\cite{Aaij:2014pli} and the CMS~\cite{Chatrchyan:2013cda} measurements. Thus we may need new SUSY models to explain this large $q^2$-discrepancy, or we need more measurements to confirm those data in large $q^2$-region.

\item As shown by Fig.\ref{Fig:AI1}, the flip of sign for the Wilson coefficients $C^{\rm eff}_{7,8}$ also makes the isospin asymmetry of MSSM MFV-II a little different from the SM prediction in low $q^2$-region. The LHCb data prefers a positive isospin asymmetry for $q^2\to 0$ which could be explained by the SM and the MSSM MFV-I models. However the LHCb data is still of large errors, thus at present, we can not draw definite conclusions on which scenario is preferable via using the present isospin asymmetry data.
\end{itemize}

Thus, we think the forward-backward and the isospin asymmetries of the $B \to K^* \mu^+ \mu^-$ decay are interesting observables to probe possible new physics beyond the SM. More accurate data, especially those in low $q^2$-region, at the LHCb or the future super $B$-factory are important for clarifying this point.

\vspace{0.5cm}

{\bf Acknowledgments}: This work was supported in part by the Natural Science Foundation of China under Grant No.11647112, No.11647113, No.11625520, No.11765007, and No.11705034; by the Project of Guizhou Provincial Department of Science and Technology under Grant No.[2017]1089; by the Project for Young Talents Growth of Guizhou Provincial Department of Education under Grant No.KY[2016]156; the Key Project for Innovation Research Groups of Guizhou Provincial Department of Education under Grant No.KY[2016]028 and the Project of Guizhou Minzu University under Grant No.16yjrcxm003.

\end{document}